\RequirePackage[hyphens]{url}
\documentclass[12pt]{article}
\usepackage[margin=1in]{geometry}
\usepackage{amsmath}
\usepackage{amsthm}
\usepackage{graphicx}
\usepackage[official]{eurosym}
\usepackage{url}
\usepackage{tablefootnote}
\usepackage{multirow}
\usepackage{subfig}
\usepackage{float}
\usepackage[hidelinks]{hyperref}
\usepackage{url}
\usepackage{breakurl}
\hypersetup{
    colorlinks=true,
    linkcolor=blue,
    filecolor=magenta,      
    urlcolor=cyan,
}
\usepackage{caption}
\usepackage{booktabs}
\usepackage{float}
\providecommand{\keywords}[1]{\textbf{keywords ---} #1}
\usepackage{tikz}
\usetikzlibrary{automata} 

\usepackage{graphics}

\title{Forecasting Mortality in the Middle-Aged and Older Population of England: A 1D-CNN Approach}

\begin{document}

\author{Marjan Qazvini \footnote{Correspondence to: Marjan Qazvini, marjan.qazvini@gmail.com} (August, 2023)}
\date{}

\maketitle

\begin{abstract}
Convolutional Neural Networks (CNNs) are proven to be effective when data are homogeneous such as images, or when there is a relationship between consecutive data such as time series data. Although CNNs are not famous for tabular data, we show that we can use them in longitudinal data, where individuals' information is recorded over a period and therefore there is a relationship between them. This study considers the English Longitudinal Study of Ageing (ELSA) survey, conducted every two years. We use one-dimensional convolutional neural networks (1D-CNNs) to forecast mortality using socio-demographics, diseases, mobility impairment, Activities of Daily Living (ADLs), Instrumental Activities of Daily Living (IADLs), and lifestyle factors. As our dataset is highly imbalanced, we try different over and undersampling methods and find that over-representing the small class improves the results. We also try our model with different activation functions. Our results show that swish nonlinearity outperforms other functions.   

\end{abstract}
\section{Introduction}
Longitudinal surveys are follow-up studies, in which participants' information is recorded at different time steps, say every two years. The difference between time series analysis and longitudinal studies is that in the former we have records of, say economic variables over a long period, whereas in the latter, we have a few records of participants, and the number of records depends on the number of participants. One problem with such studies is that we may have a large number of drop-outs, known as right-censoring in survival analysis, due to death, illness, immigration, etc. Another difference is that in time-series analysis, we aim to predict, say future stock prices based on past data, whereas in longitudinal study our goal is to predict a target based on some features. Longitudinal studies are used to study life events such as clinical psychology. Generalised linear models (GLMs) (McCullagh and Nelder, 1989), and generalised linear mixed models (GLMMs) (Frees, 2004) are traditional methods used in longitudinal studies. Qazvini (2023) employs GLMM to analyse the survival rate among the English population using the ELSA dataset. However, in recent years, some research use machine learning (ML) models and algorithms to compare their predictability with traditional regression models. For example, XGBoost is used to analyse the Household, Income, and Labour Dynamics in Australia (HILDA) survey over 13 years (Sheetal et al., 2023), random forest (RF) (Breiman, 2001) is used to predict cognitive impairment among the participants of China Health and Retirement Longitudinal Survey (CHARLS) (Liu et al., 2023). In this research, RF is found to have a higher predictive power than a logistic regression based on the area under the receiver operating characteristic curves. Catboost (Dorogush et al., 2017), XGboost (Chen et al. 2016), Lightgbm (Guolin et al. 2017), random forest, gradient boosting classifier, and logistic regressions are used by (Silva et al. 2023) to predict mortality among patients, suffering from cancers such as clone, stomach, breast, prostate, etc.  Recurrent neural networks (RNNs) (De Brouwer et al. 2019), bayesian probabilistic tensor factorisation (BPTF) (Simm et al. 2017) are also used to predict disability progression among MS patients (De Brouwer et al. 2021).

\par In this study, we consider the English Longitudinal Study of Ageing (ELSA). This is a survey conducted every two years and targeted the English population aged 50 and over. It is a collection of economic, social, psychological, cognitive, health, biological, and genetic data. This large survey has been extensively studied by many researchers using traditional methods see e.g. Demakakos et al, 2018, Pongiglione et al, 2016, Qazvini 2023, and the references therein. In recent years, the applications of ML models to this dataset have drawn the attention of researchers in different areas such as the prediction of cholesterol among participants (Fazakis et al. 2021, and Dritsas et al, 2021), prediction of frailty (De Cunha Leme et al. 2023), and the risk of dementia (Stamate et al. 2022). Our aim is to use one-dimensional Convolutional Neural Network (1D-CNN) to predict mortality in wave 6. Most machine learning models, applied to longitudinal data, are based on ensemble trees such as RF and XGboost. Therefore, to the best of our knowledge, this is the first time that 1D-CNN is used in longitudinal data. The rest of this paper is organised as follows: Section 2 describes ELSA dataset. Section 3 explains 1D-CNN. Section 4 discusses the results and Section 5 concludes.

\begin{table}
\center
\footnotesize
\caption{The number of participants in each wave. NatCen: Technical Report, wave 6}
\label{tabw}
\begin{tabular}{|ccccc|}
\hline
Waves		&~	Cohort 1	&~	Cohort 3	&~ Cohort 4	&~	End of life interviews  \\\hline
1) 2002-2003    &~	11,391	&~		-	&~	-		&~	-	\\
2) 2004-2005	&~	8,780	&~		-	&~	-		&~     133	\\
3) 2006-2007	&~	7,535	&~	1,275	&~	-		&~	369	\\
4) 2008-2009	&~	6,623	&~	972		&~	2,291	&~	234	\\
5) 2010-2011	&~	6,242	&~	936		&~	1,912	&~	-	\\
6) 2012-2013	&~	-		&~	-		&~	-		&~	240	\\
\hline
\end{tabular}
\end{table}

\section{The description of data}
The English Longitudinal Study of Ageing (ELSA) commenced in 2002. The first cohort was selected from respondents to the Health Survey for England (HSE) in 1998, 1999, and 2001 and included people born on or before February 29, 1952, i.e., aged 50 and older. The first ELSA wave was in 2002-2003. To make sure ELSA is designed to be representative of people aged 50 and over in England, in waves 3 and 4, a refreshment cohort of people just entering their 50s was introduced. In waves 2, 3, 4, and 6 an End of Life interview was conducted with the purpose of finding out about the health and socio-economic situations of people just before their death (Steptoe et al, 2013, and Blake et al. 2015). 
Table \ref{tabw} shows the number of participants in each wave from different cohorts and the number of deaths. In this study, our goal is to predict the status of participants after 12 years, i.e. 6 waves. We only consider participants from cohort 1, who participate in all 5 waves. We use questions, which are asked in all 5 waves and related to socio-demographics such as age, gender, marital and employment status, health, and lifestyles as our features, and the status of individuals in wave 6 as our target feature. We only consider features, with less than $3\%$ missing values in wave 1. Most of our features are categorical except age, which is numerical. We fill in missing values using the last and the next available observations, such that missing values in wave 1 are filled in by wave 2, wave 2 by wave 1, wave 3 by wave 4, and wave 4 by wave 5.

\begin{table}
\center
\footnotesize
\caption{Longitudinal data in short format}
\label{tabs}
\begin{tabular}{|lccccccccccccccc|}
\hline
		&& \multicolumn{13}{c}{Features}																& Target\\\hline
		&&	age1		&& age2	 & & age3   && ~sex1   & sex2        & sex3       & &~ illness   & illness       & illness 	& $y$ \\\hline
$P_{1}$ 	&&	56		&&	58	 & &	60	   &&	~1	    &	1	      &		1	&&~	0	&	1	   &		1	& 0	  \\\hline
$P_{2}$	&&	62		&&	64	 & &	66	   && ~0	    &	0	      &		0	&&~	0	& 0		   &		1	&	0	\\\hline
$P_{3}$	&& 	70		&&	72	 & &	74	   &&	~1	    &	1	      &		1	&&~	1	&	1	   &		1	&	1\\
\hline
\end{tabular}
\end{table}

\begin{table}
\center
\footnotesize
\caption{Longitudinal data in long format}
\label{tabl}
\begin{tabular}{|lcccccccccc|}
\hline
			&&& \multicolumn{7}{c}{Features}		& Target\\\hline
			&&&		age	&&~ sex	&&~ illness&&& $y$\\\hline
$P_{11}~$		&&& 		56	&&~	1	&&~	0	&&&	 0\\
$P_{12}~$	&&&		58	&&~	1	&&~	1	&&&	0\\
$P_{13}~$	&&&		60	&&~	1	&&~	1	&&&	0\\\hline
$P_{21}~$	&&&		62	&&~	0	&&~	0	&&&	 0\\
$P_{22}~$	&&& 		64	&&~	0	&&~	0	&&&	0\\
$P_{23}~$	&&&		66	&&~	0	&&~	1	&&&	 0\\\hline
$P_{31}~$	&&& 		70	&&~	1	&&~	1	&&&	0\\
$P_{32}~$	&&&		72	&&~	1	&&~	1	&&&	0\\
$P_{33}~$	&&&		74	&&~	1	&&~	1	&&&	1\\
\hline
\end{tabular}
\end{table}

\noindent There are two ways to present longitudinal data in a table: short format, where the rows are unique participants and columns are all features over the waves, and long format, where rows contain repeated observations for participants. Tables \ref{tabs} and \ref{tabl} illustrate these two formats for 3 participants $(P_!, P_2, P_3)$, 3 features (age, sex, illness),  3 waves, and 1 target $y$. In Table \ref{tabs}, $P_{1}$ represents individual 1 and age1 represents age in wave 1, \textit{age2} represents age in wave 2, and so on. In Table \ref{tabl} $P_{11}$ represents participant $1$ in wave 1, and $P_{12}$ represents participant 1 in wave 2. In this table, columns represent features across all waves. In both tables, $y$ represents individuals' status, where $1$ denotes death and $0$, survival. Here, we use the short format. In this case, in each row, we have unique participants with features for consecutive waves next to each other. Therefore, there is a relationship between consecutive features, which is one of the requirements of CNN that we discuss in the next section. This is different from traditional modelling, such as GLMMs, where longitudinal data are usually presented in a long format. After extracting information for participants who have records for 5 waves, we have $5,314$ unique individuals (rows) and 52 features for 5 waves, i.e. 260 columns.

\section{One dimensional convolutional neural networks (1D-CNNs)}
CNNs are effective when data are homogeneous or there is a relationship between features. Tabular data are not homogeneous, and features are not necessarily in particular order. However, in some cases, like longitudinal data, we can see the evolution of features through time. If we write longitudinal data in a short format, we can see that at least some features have to be in a particular order. We use 2D-CNNs with images that have 3 dimensions. Tabular data have 2 dimensions, and we can use 1D-CNNs. The idea of 1D-CNNs originates from a Kaggle competition \footnote{\url{https://www.kaggle.com/competitions/lish-moa/discussion/202256\#1106810}. Last accessed July 2023} (Ye et al. 2023). In our case, we have features from 5 waves. Therefore, every 5 cells must be arranged in a particular order.  
Let $x$ denote our input with features $a, b, c$, and $d$. Further, let $w$ be the kernel (filter). We can then slide the filter over the sequence in $x$. The output is then called a feature map. In our case, we use a stride of 5, meaning that after sliding over the first 5 features, the filter jumps to the next 5 features. In other words, we have segments of 5 cells that are independent of each other. Therefore, in the case of the following $x$ and $w$, the feature map is an array of 4 with the values, given by

\begin{equation}
f_1 = \sum_{i=1}^5 a_i w_i ; ~~~ f_2 = \sum_{i=1}^5 b_i w_i ;~~~
f_3 = \sum_{i=1}^5 c_i w_i ; ~~~ f_4 = \sum_{i=1}^5 d_i w_i .
\end{equation}

\begin{table}[h]
\footnotesize
\label{tabi}
\begin{tabular}{|c|c|c|c|c||c|c|c|c|c||c|c|c|c|c||c|c|c|c|c|}
\multicolumn{20}{l}{Input: $x$}\\
\hline
$a_1$~&$a_2$~&$a_3$~&$a_4$~&$a_5$~&$b_1$~&$b_2$~&$b_3$~&$b_4$~&$b_5$~&$c_1$~&$c_2$~&$c_3$~&$c_4$~&$c_5$~&$d_1$~&$d_2$~&$d_3$~&$d_4$~&$d_5$~\\
\hline
\end{tabular}
\end{table}

\begin{table}[h]
\footnotesize
\label{tabi}
\begin{tabular}{|c|c|c|c|c|}
\multicolumn{5}{l}{Kernel: $w$}\\
\hline
$w_1$~&$w_2$~&$w_3$~&$w_4$~&$w_5$~\\
\hline
\end{tabular}
\end{table}

\begin{table}[h]
\footnotesize
\label{tabi}
\begin{tabular}{|c|c|c|c|}
\multicolumn{4}{l}{Feature map}\\
\hline
$f_1$~&$f_2$~&$f_3$~&$f_4$~\\
\hline
\end{tabular}
\end{table}

\section{Results}
Our dataset is highly imbalanced. Only $1\%$ of individuals participate in 5 waves and die between the 5th and the 6th wave. Therefore, we need to create synthetic data. First, we divide our dataset into $80\%$ training set and $20\%$ validation set and then divide our training set into $80\%$ training set and $20\%$ test set. We stratify our target feature to make sure that the percentage of the event of interest, i.e. death is the same in all three sets. We then use our training set and try 5 methods to create synthetic data, namely, random oversampling (ROS) (Menardi and Torelli, 2014), synthetic minority oversampling technique (SMOTE) (Chawla et al. 2002), adaptive synthetic sampling (ADASYN) (He et al, 2008), SMOTE with edited nearest neighbour (SMOTEEN) (Batista et al. 2003), and SMOTETomek (Batista et al, 2004), which is a combination of SMOTE and a modification of nearest neighbour introduced by (Tomek 1976). The last two methods are hybrid methods that over-represent the under-represented class and under-represent the over-represented class. After adding synthetic data to our training set, we normalise age. Then we design the architecture of our network such that different features are not muddled. Hence, the first layer after the input layer is a 1D-CNN with 8 filters of size 1 and a stride of 1. In this layer, we just try to add new weights to our model. In the 2nd layer, we use a 1D-CNN with 16 filters of size 5 with a stride of 5 to downsize our data. After that, we use a flatten layer and finally, the output layer is a dense layer with 1 node and a sigmoid activation function. We try different activation

\begin{figure}
\includegraphics[scale=0.25]{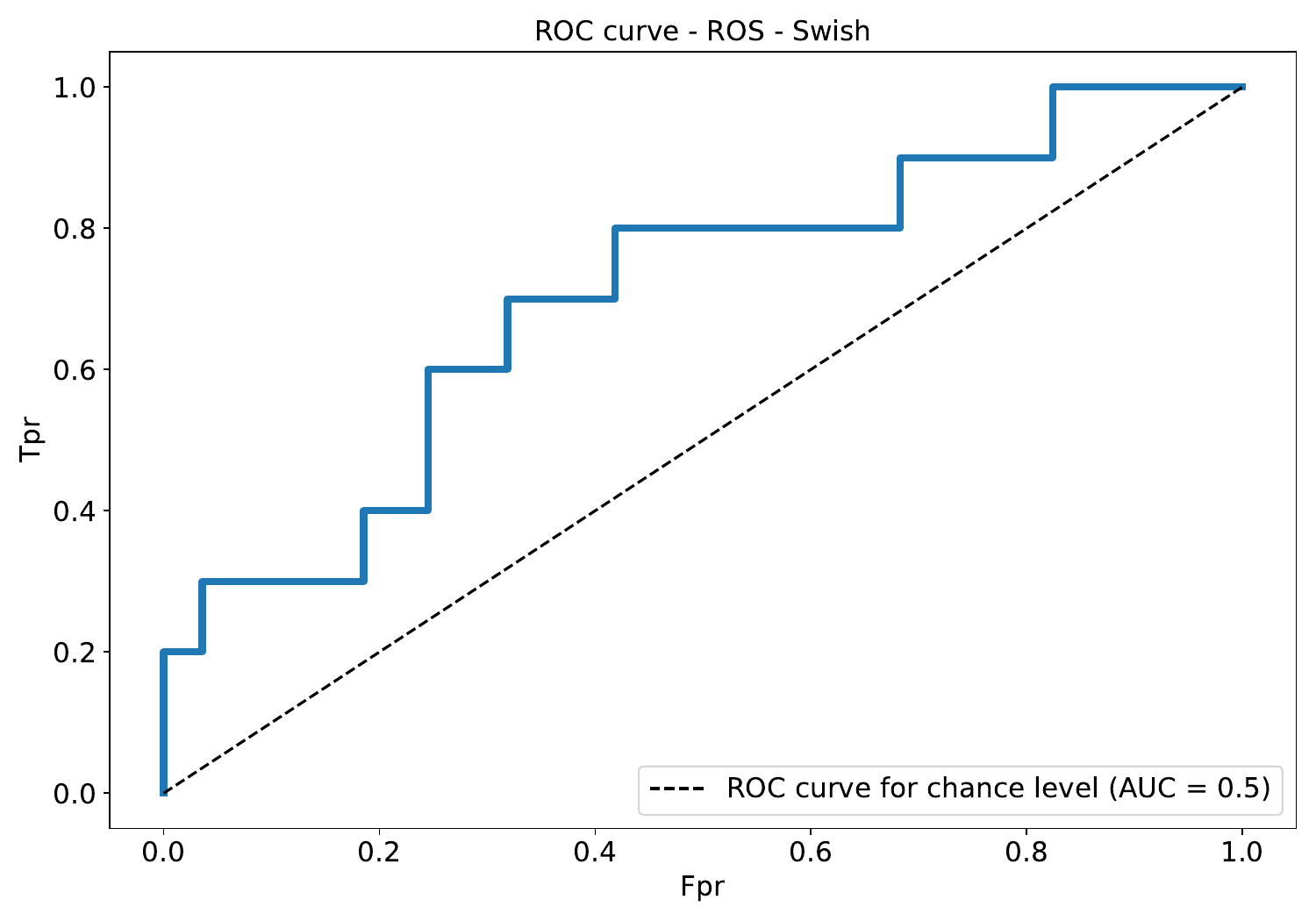}
\includegraphics[scale=0.25]{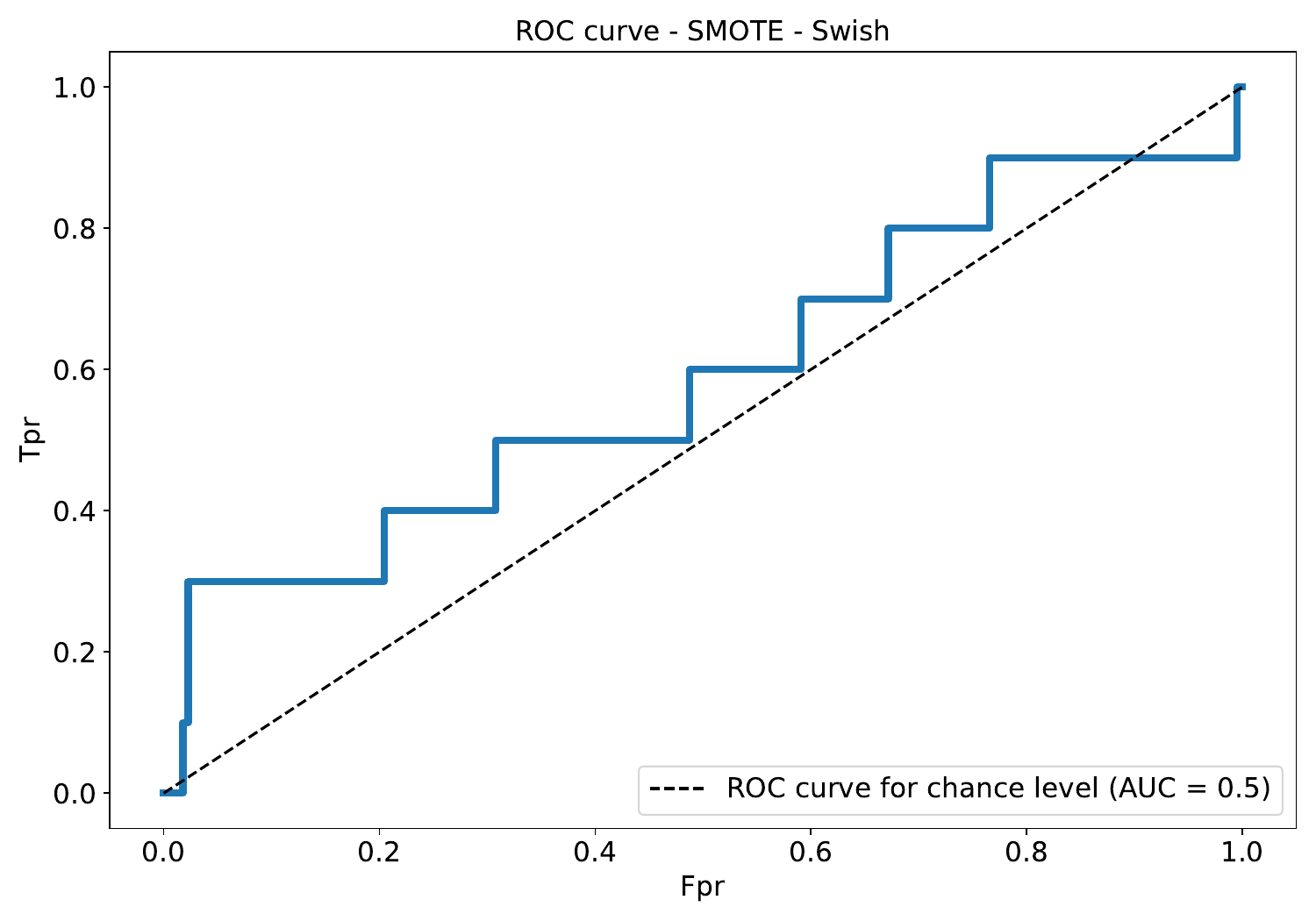}
\includegraphics[scale=0.25]{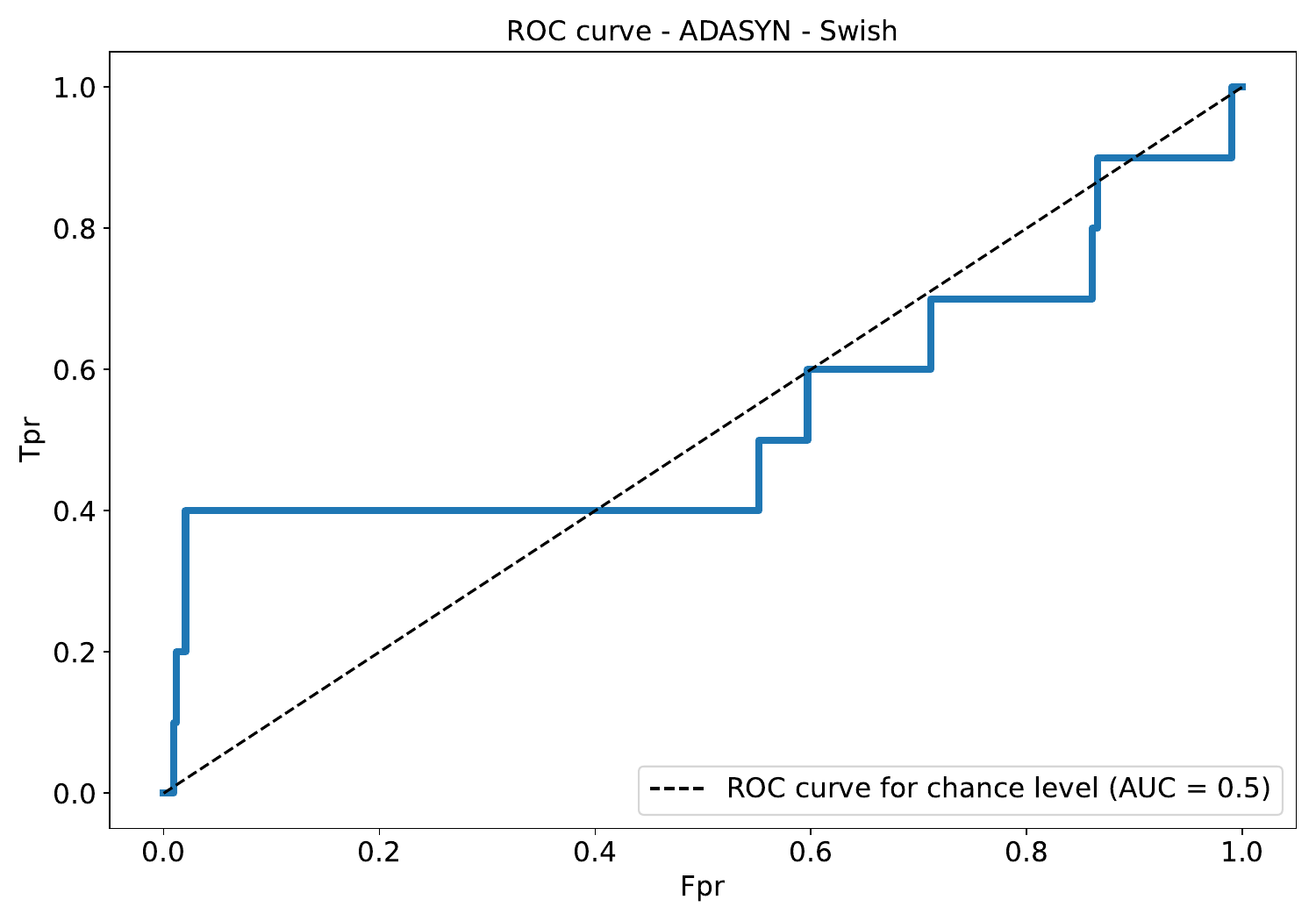}
\includegraphics[scale=0.25]{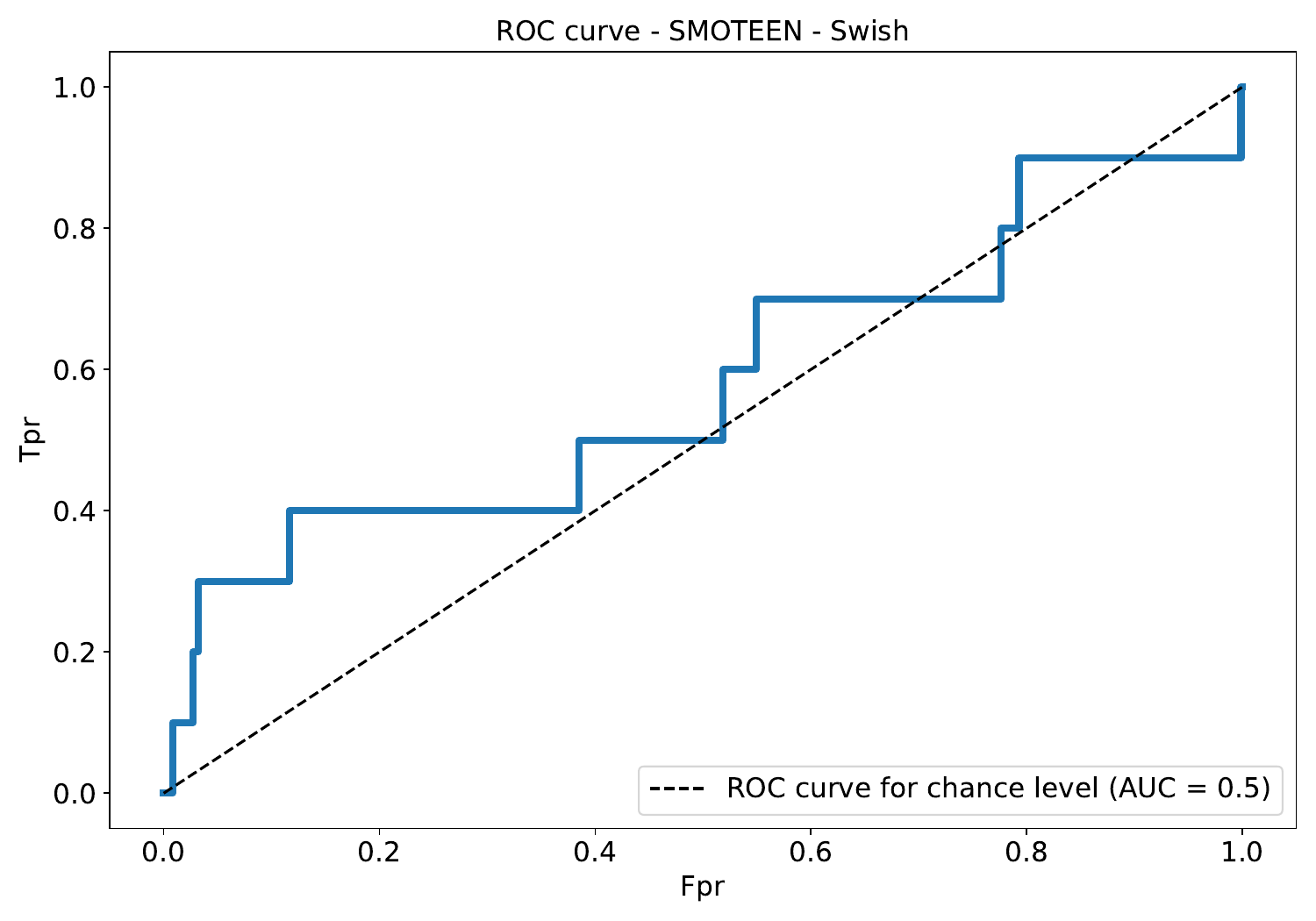}
\includegraphics[scale=0.25]{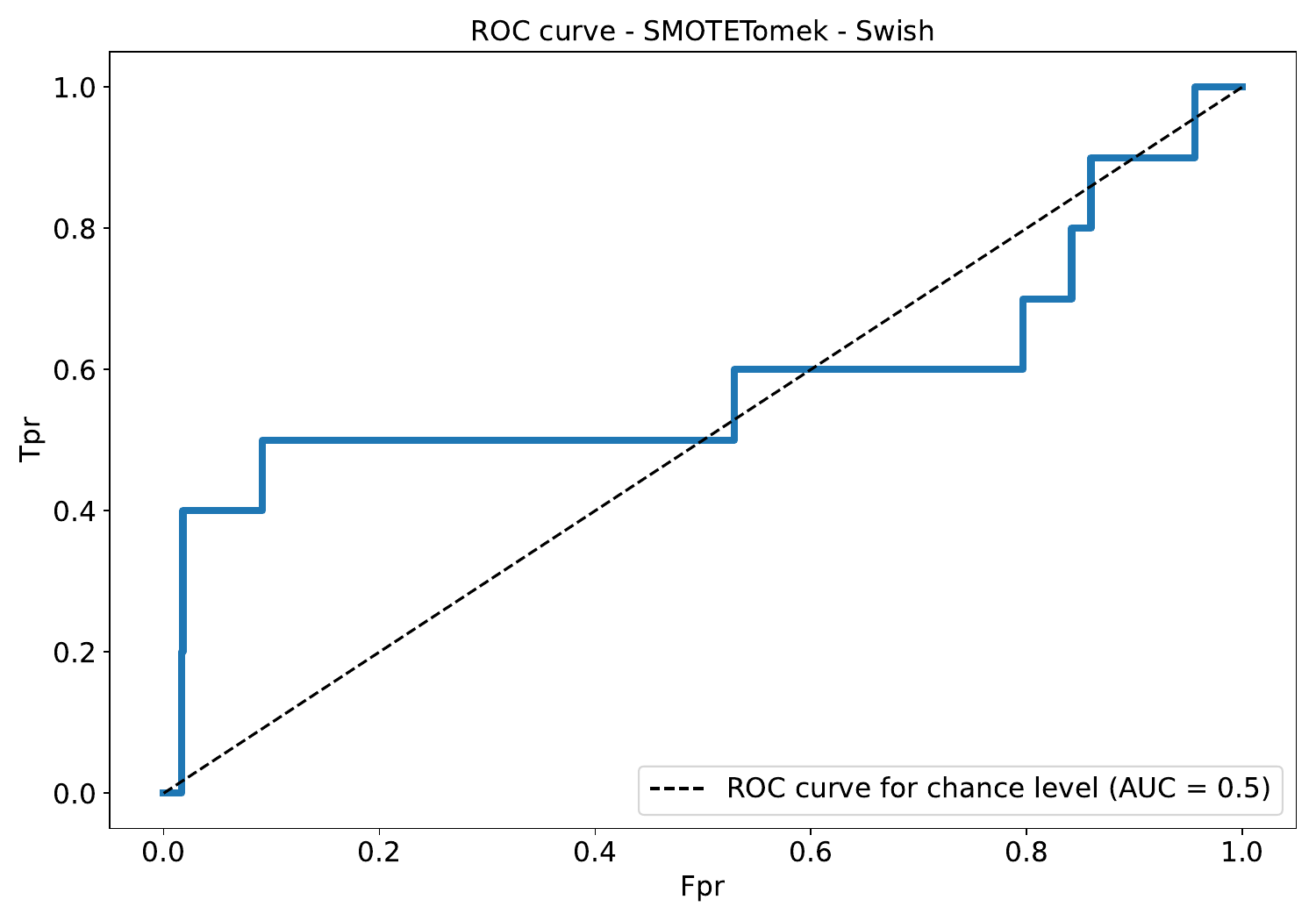}
\caption{ROC curves for swish nonlinearity and different over/undersampling methods}
\label{figauc}
\end{figure}

\begin{table}
\center
\footnotesize
\caption{Oversampling/undersampling methods}
\label{tabr}
\begin{tabular}{|l|ccc||ccc||ccc|}
\hline
		 &	 \multicolumn{3}{c||}{ROS}&  \multicolumn{3}{c||}{SMOTE}	&\multicolumn{3}{c|}{ADASYN}		\\	\hline
	         &Loss & Accuracy    &    AUC   &  Loss &Accuracy  	&AUC	&Loss   &	Accuracy	        & AUC	\\\hline
ReLU        &0.6124& 0.9835     &  0.5369&0.4786&0.9741 	&0.4326	&0.5141& 0.9718		&0.4527		\\
SeLU	 &0.6631& 0.9788     & 0.6799&0.4840&0.9553        &0.6064	&0.3600&0.9741		&0.6239		\\
ELU	         &0.5417& 0.9788     & 0.6435&0.4174&0.9600        &0.5994	&0.3787&0.9777		&0.6048		\\
Swish        &0.4121& 0.9859     & 0.7045&0.3526&0.9694        &0.5913	&0.3599&0.9718		&0.5361		\\
Leaky        &0.6767& 0.9741     &0.5957&0.4775&0.9741         &0.4492	&0.4404&0.9777		&0.4338		\\
\hline
\end{tabular}
\end{table}

\noindent functions in hidden layers including ReLU (Nair and Hinto, 2010), SeLU (Klambauer et al, 2017), ELU (Clevert et al. 2015), Swish (Prajit et al. 2017), and leaky ReLU (LReLU) (Maas et al. 2013). We use Adam optimiser (Kingma and Ba, 2015) with learning rate $0.01$ and binary cross entropy loss function, and apply different metrics such as loss, accuracy, and area under the receiver operating characteristic (ROC) curve, i.e. (AUC) to compare our model with different nonlinearities and different over/undersampling methods. We use early-stopping callbacks and save the best model. Then, we try the best model on the test set. The results are provided in Tables \ref{tabr} and \ref{tabrh}. As we can see the hybrid model does not give rise to better results. In terms of loss, ROS produces the worst results. SMOTE and ADASYN result in the smallest losses. These two are oversampling methods. Therefore, based on our findings, over-representing the under-represented class works better than the combinations of over and undersampling methods. Comparing different nonlinearities, we can see that ELU and swish perform better than other activation functions. In general, in terms of AUC, none of the functions perform well. This is usually the problem with longitudinal data. Although it is possible to train a well-fitted model, it is difficult to find a model that can generalise well to an unseen dataset. Fig \ref{figauc} illustrates ROC curves for the swish activation function and different over and undersampling methods. As we can observe in terms of AUC, the best result is obtained with the ROS method.

\begin{table}
\center
\footnotesize
\caption{Combination of oversampling/undersampling methods }
\label{tabrh}
\begin{tabular}{|l|ccc||ccc|}
\hline
		 ~&	 \multicolumn{3}{c||}{SMOTEEN}&  \multicolumn{3}{c|}{SMOTETomek}	\\\hline
	         ~&~	Loss 	&~ Accuracy  &	~AUC  	~&~Loss   &~Accuracy  &~ AUC		\\\hline
ReLU        ~&~	0.5784	&~ 0.9683	     &	~0.5128	~&~0.5501&~0.9706	     &~0.5188		\\
SeLU	 ~&~	0.4826	&~ 0.9741	     &~0.6304	~&~0.4004&~0.9753	     &~0.6467		\\
ELU	         ~&~ 0.5521	&~ 0.9671	     &~0.6156	~&~0.4132&~0.9694	     &~0.6206		\\
Swish        ~&~	0.4482	&~ 0.9777	     &~0.5794	~&~0.3626&~0.9753	     &~0.5856		\\
Leaky        ~&~	0.5567	&~ 0.9624	     &~0.5996	~&~0.4003&~0.9718	     &~0.4580		\\
\hline
\end{tabular}
\end{table}

\section{Conclusion}
In this study, we used English Longitudinal Study of Ageing (ELSA) to predict participants' status after 6 waves. We only considered those individuals who participated in all 5 waves. 
Our dataset was highly imbalanced and we needed to create synthetic data to train a model. We used different under and oversampling methods and trained a 1D-CNN model. We try different activation functions and found that swish outperformed the other nonlinearities. We also found that oversampling the under-represented class gives rise to better results than the hybrid methods.

\end{document}